\documentclass[final,3p,times,twocolumn]{elsarticle}

\usepackage{amssymb}
\journal{Results in Physics}

\begin{document}

\begin{frontmatter}

%% Title, authors and addresses

%% use the tnoteref command within \title for footnotes;
%% use the tnotetext command for theassociated footnote;
%% use the fnref command within \author or \address for footnotes;
%% use the fntext command for theassociated footnote;
%% use the corref command within \author for corresponding author footnotes;
%% use the cortext command for theassociated footnote;
%% use the ead command for the email address,
%% and the form \ead[url] for the home page:
%% \title{Title\tnoteref{label1}}
%% \tnotetext[label1]{}
%% \author{Name\corref{cor1}\fnref{label2}}
%% \ead{email address}
%% \ead[url]{home page}
%% \fntext[label2]{}
%% \cortext[cor1]{}
%% \affiliation{organization={},
%%             addressline={},
%%             city={},
%%             postcode={},
%%             state={},
%%             country={}}
%% \fntext[label3]{}

\title{Bound state of a $^3$He atom at the interface of crystal and superfluid $^4$He}

%% use optional labels to link authors explicitly to addresses:
%% \author[label1,label2]{}
%% \affiliation[label1]{organization={},
%%             addressline={},
%%             city={},
%%             postcode={},
%%             state={},
%%             country={}}
%%
%% \affiliation[label2]{organization={},
%%             addressline={},
%%             city={},
%%             postcode={},
%%             state={},
%%             country={}}

\author[inst1]{Massimo Boninsegni}

\affiliation[inst1]{organization={Department of Physics},%Department and Organization
            addressline={University of Alberta}, 
            city={Edmonton},
            postcode={T6G 2E1}, 
            state={Alberta},
            country={Canada}}

%\author[inst2]{Author Two}
%\author[inst1,inst2]{Author Three}

%\affiliation[inst1]{organization={Department Two},%Department and Organization
%            addressline={Address Two}, 
%             city={City Two},
%            postcode={22222}, 
%            state={State Two},
%            country={Country Two}}

\begin{abstract}
%% Text of abstract
%The possible suppression of crystallization in
%liquid mixtures of parahydrogen (\ph2) and
%orthodeuterium (\od2) at temperatures below the
%crystallization temperature (13.8 K) isWe study, by quantum Monte Carlo simulations, the ground state of a harmonically confined dipolar Bose
The bound state of a $^3$He atom at the interface between coexisting crystalline and superfluid phases of  $^4$He is studied theoretically by means of first principle Quantum Monte Carlo simulations. We consider both the case of a solid/superfluid $^4$He interface, as well as that of a few solid layers of $^4$He forming on an attractive substrate. 
It is observed that the $^3$He bound state is sharply localized in a well-defined, quasi-2D layer of $^4$He in the region intermediate between the solid and the liquid. The layer in which the $^3$He atom resides displays an intriguing interplay of atomic localization and quantum-mechanical exchanges; is not necessarily the first ``non-solid'' layer, as suggested in previous studies, and it is affected by the attractive strength of the substrate.

\end{abstract}

%%Graphical abstract
%\begin{graphicalabstract}
%\includegraphics{grabs}
%\end{graphicalabstract}

%%Research highlights
%\begin{highlights}
%\item Research highlight 1
%\item Research highlight 2
%\end{highlights}

%\begin{keyword}
%% keywords here, in the form: keyword \sep keyword
%keyword one \sep keyword two
%% PACS codes here, in the form: \PACS code \sep code
%\PACS 0000 \sep 1111
%% MSC codes here, in the form: \MSC code \sep code
%% or \MSC[2008] code \sep code (2000 is the default)
%\MSC 0000 \sep 1111
%\end{keyword}

\end{frontmatter}

%% \linenumbers

%% main text
\section{Introduction}\label{intro}
The existence of a bound state of a $^3$He atom either at a free liquid $^4$He surface, or at the interface between crystalline and superfluid phases, is a subject of longstanding interest, due to its relevance to different aspects of the phenomenology of superfluid $^4$He, as well as for the intriguing possibility of stabilizing a quasi-2D $^3$He gas with novel superfluid properties \cite{edwards1978,bashkin1980,miyake1983}. 
\\ \indent 
It was first proposed by Andreev \cite{andreev1966} that localized states of a $^3$He atoms at the surface of superfluid $^4$He may exist, as an {\em ad hoc} mechanism to account for observed behaviour of the surface tension of isotopic helium mixtures, markedly different from that of bulk $^4$He \cite{atkins1965}. The first attempt at providing a microscopic explanation of such bound states, whose occurrence is rather counter-intuitive, is the variational theory of Lekner \cite{lekner1970}.
Later on, Pavloff and Treiner \cite{pavloff1991,treiner1993} contended that the same physical mechanism described by Lekner's theory could also lead to the formation of $^3$He bound states in the first non-solid (i.e., superfluid) layer of  $^4$He near a substrate, including at the interface between coexisting superfluid and solid phases of $^4$He. Using a ground state density functional (DFT) approach, an estimate of the $^3$He binding energy close to 5.7 K (with respect to vacuum) was obtained, along with one for the $^3$He effective mass, approximately 2.3 times the bare mass \cite{treiner1993}. 
\\ \indent 
On the one hand, the existence of a bound state is the obvious consequence of the attraction experienced by the $^3$He atom, which moves through superfluid $^4$He as an essentially free particle \cite{landau1948}, toward the denser phase. Clearly, however, in order to address quantitative issues such as the localization, binding energy, and the proximity of such a $^3$He bound state to the $^4$He crystal, which are crucially relevant in the interpretation of a number of experiments (especially in nanoscale size confinement), a calculation based on a reliable model of the interface, its roughness, and the interplay of superfluidity and localization within it, is required. Over three decades after the first such calculation, no  independent microscopic study has been carried out, possibly overcoming (some of) the limitations of that of  Ref. \cite{treiner1993}, mainly its reliance on a heuristic density functional. It is worth mentioning that the experimental search for $^3$He substrate states has yielded so far conflicting evidence, and sizeable uncertainty concerning the binding energy \cite{carmi1988,wang1992,ketola1993,draisma1994,sheldon1995,ross1995,rolley1995,chang2021}.  
\\ \indent 
We revisit this problem in this paper, reporting the results of a first principle theoretical study, based on Quantum Monte Carlo (QMC) computer simulations, of the bound state of a single $^3$He atom at the interface between coexisting crystal and superfluid phases, at a pressure of 25 bars. We adopt some of the basic assumptions built in the calculation of  Ref. \cite{treiner1993}, in that we consider both bulk solid and liquid phases, far away from where the $^3$He atom is located, as uniform, continuous media; on the other hand, in the (interfacial) region of interest all atoms are modeled explicitly, and a realistic interatomic potential for helium is utilized. This allows one to study layering of atoms near the solid substrate, and the ensuing emergence of superfluidity as one moves away from it, both believed to affect crucially the nature of the $^3$He bound state \cite{lekner1970,pavloff1991,treiner1993}, in a way does not require any {\em a priori} physical assumption. 
\\ \indent
We consider two cases, namely that of the interface separating coexisting solid and superfluid phases, as well as that in which solid layers of $^4$He form on top of an attractive substrate.
What is observed in both cases is that the bound state of the $^3$He atom is sharply localized in the interfacial region, whose width is roughly 20 \AA. The $^3$He is confined within a fairly high density, quasi-2D $^4$He layer. The physical character of this layer scarcely fits either the ``solid'' or ``liquid'' definition, displaying qualities of both phases.  We compute the $^3$He effective mass, and in both cases we obtain a value above three times the bare mass, significantly above the estimate (2.3) of Ref. \cite{treiner1993} and consistent with the high density of the $^4$He environment experienced by the $^3$He atom.
\\ \indent
The remainder of this paper is organized as follows: in section \ref{mod} we describe the microscopic model of the system; in Sec. \ref{meth} we briefly describe our methodology; we present and discuss our results in Sec. \ref{res} and finally outline our conclusions in Sec. \ref{conc}.

\section{Model}\label{mod}
\begin{figure}[h]
\centering
\includegraphics[width=\linewidth]{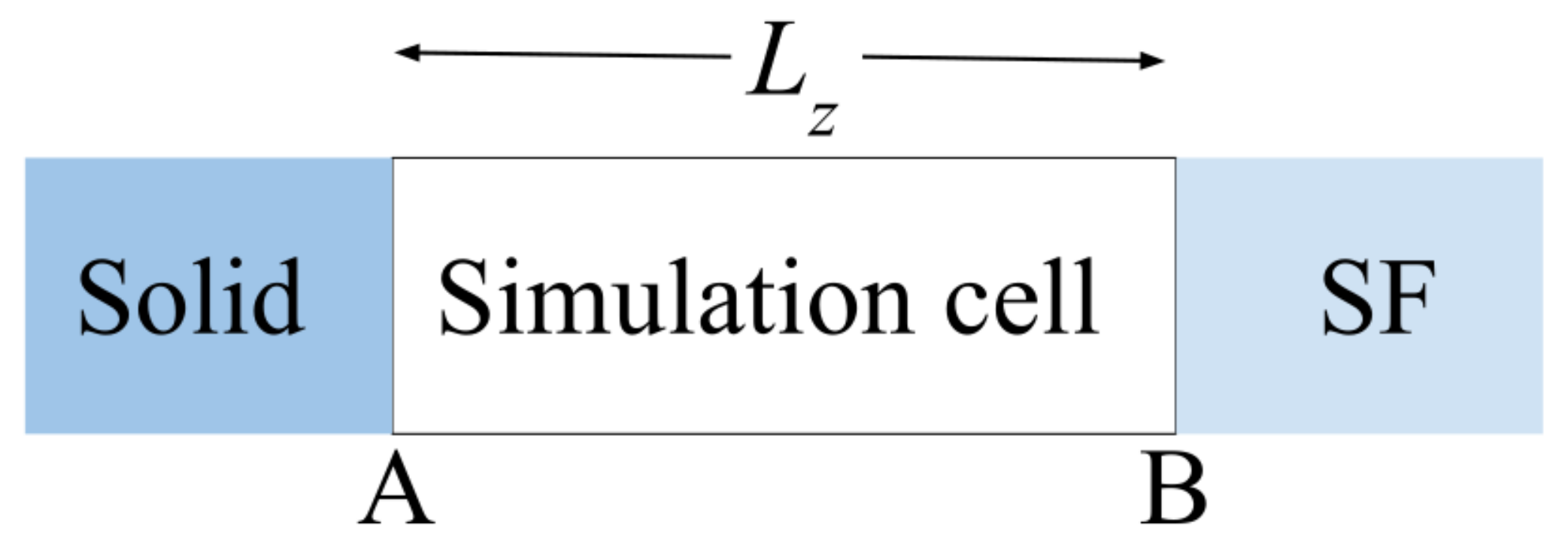}
\caption{Schematic of the simulation setup. The simulation  cell is elongated, sandwiched between homogeneous solid and superfluid phases,
modeled as homogeneous, continuous media.}
\label{scheme}
\end{figure}

The model system simulated here comprises $N=864$ He atoms, one of them being of the light isotope $^3$He and all others of $^4$He, enclosed in a parallelepipedal cell  of sizes $L\times L\times L_z$, with $L=20.075$ \AA\ and $L_z=80.8$ \AA. Periodic boundary conditions are utilized in the $x$ and $y$ directions, not in the long ($z$) direction, as the simulation cell ends on both sides with two square walls (A and B in Fig. \ref{scheme}) representing two different substrates\footnote{This setup was first proposed in Ref. \cite{khairallah2006}}.

One of the two substrates (A) is that on which crystalline layers of $^4$He form, the other (B) can be regarded as the surface of superfluid $^4$He at freezing density. We consider here two distinct situations, namely one in which the phase to the left of wall A is solid $^4$He at the melting density (i.e., one is looking at the physical interface between solid and superfluid phases of $^4$He),  the other in which it is a generic, homogeneous  medium, more strongly attractive than $^4$He, as a result of which a number of solid layers of $^4$He form on the substrate.

The interaction of the He atoms in the simulation cell with the superfluid and solid beyond the substrates is accounted for by means of effective interactions of the He atoms with the substrates, described by suitably parametrized ``3-9'' potentials
\begin{equation}\label{39}
    V(z) =\frac{D}{2}\ \ \biggl [ \biggl ( \frac {a}{z}\biggr )^9-3\ \biggl (\frac{a}{z}\biggr )^3\biggr ]\ \ ,
\end{equation}
where $z$ is the distance of the atom from the substrate, $D$ is the depth of the attractive well of the interaction, and $a$ can be seen as the distance of closest approach of an atom to the substrate, below which the atom experiences a strong (hard core) repulsion. The values of $D$ and $a$ are chosen  to represents the specific media considered here. We have altogether three different parameter sets, corresponding to solid (superfluid) $^4$He at melting (freezing) density\footnote{The parameters $D$ and $a$ are related to the density $\rho$ of the medium (i.e., solid helium or glass, or superfluid helium), represented as a continuous, homogeneous semi-infinite slab, as well as to values of $\epsilon$ and $\sigma$ of the Lennard-Jones potential describing the interaction a helium atom and an atom or molecule of the medium. Specifically, it can be easily shown that  $D\approx 2.2077\ \epsilon\rho\sigma^3$, $a\approx 0.8584\ \sigma$. }, as well as a third set with parameters that are typically used to describe the interaction of He atoms with glass \cite{pricaupenko1995,boninsegni2010}, and we shall henceforth refer to it as such. All of this is summarized in Table \ref{table1}.
\begin{table}[]
    \centering
    \begin{tabular}{c|c|c}
      Set &$D$ (K)   &$a$ (\AA)  \\
       1  & 10.75 & 2.19 \\
       2  & 9.80 & 2.19 \\
       3  & 100 & 2.05 \\
    \end{tabular}
    \caption{The different parameters set utilized to represent the interaction of He atoms in the simulation cell with the various substrates considered in this work. Set 1 (2) corresponds to the surface of solid (superfluid) He at the equilibrium $T=1$ K melting (freezing) density, namely $\rho_m=0.0287$ ($\rho_f=0.0260$) \AA$^{-3}$, from Ref. \cite{grilly1973}. Set 3 pertains to a glass substrate.}
    \label{table1}
\end{table}
The interaction between two He atoms is described by the accepted Aziz pair potential \cite{aziz1979}.

The length $L_z$ of the simulation cell was adjusted to ensure that the local $^4$He density approach $\rho_f$, on moving away from substrate A. The value of $L_z$ quoted above (80.8 \AA), was found adequate for both types of substrate A considered here (i.e., entries 1 and 3 in Table \ref{table1}).

\section{Methodology}\label {meth}
The QMC methodology adopted here is the canonical \cite{mezz,mezz2} continuous-space Worm Algorithm \cite{worm,worm2}, a finite temperature ($T$) quantum Monte Carlo (QMC) technique. Although we are clearly interested in low temperature physics, 
finite temperature  methods have several advantages over ground state ones, for investigating Bose systems (for an extensive discussion  of this subject, see for instance Ref. \cite{rmp95}); in particular, they are unaffected by the bias of existing ground state methods, arising from the use of a trial wave function, as well as from the control of a population of walkers \cite{psb,phasesep}. 

Details of the simulations carried out in this work are standard, and therefore the reader is referred to the original references. We used the fourth-order approximation (see, for instance, Ref. \cite{jltp2}), and observed convergence of all the physical estimates for a value of time step $\tau=3.125\times 10^{-3}$ K$^{-1}$.

Besides energetic and structural properties, such as integrated density profiles computed along the relevant ($z$) direction, as well as the the global $^4$He superfluid response (using the winding number estimator \cite{rmp95}),  a physically important outcome of the simulation must be a quantitative assessment of the interplay of superfluidity and local crystalline order in the interfacial region. Although there exists, within QMC, an accepted estimator of the local superfluid density \cite{paesani,mezz3,dang2008}, it is numerically rather noisy, making impractically lengthy simulations necessary in order to accumulate the required statistics.
\\ \indent 
Because superfluidity is intimately connected with exchanges of identical particles, one can gain local insight into the emergence of superfluidity by computing the frequency with which a $^4$He atom is involved in a quantum-mechanical exchange, as a function of its distance from substrate A. While {\em not} an estimator of the local superfluid density \cite{mezz3}, this quantity nonetheless is well-suited to address the physical issues of interest here, as exchanges are known to be strongly suppressed in the crystalline phase of $^4$He \cite{boninsegni2006}, .
\\ \indent
Finally, we also computed the $^3$He effective mass, using the methodology illustrated in Ref. \cite{boninsegni1995}.
\section{Results}\label{res}
\begin{figure}[t]
\centering
\includegraphics[width=\linewidth]{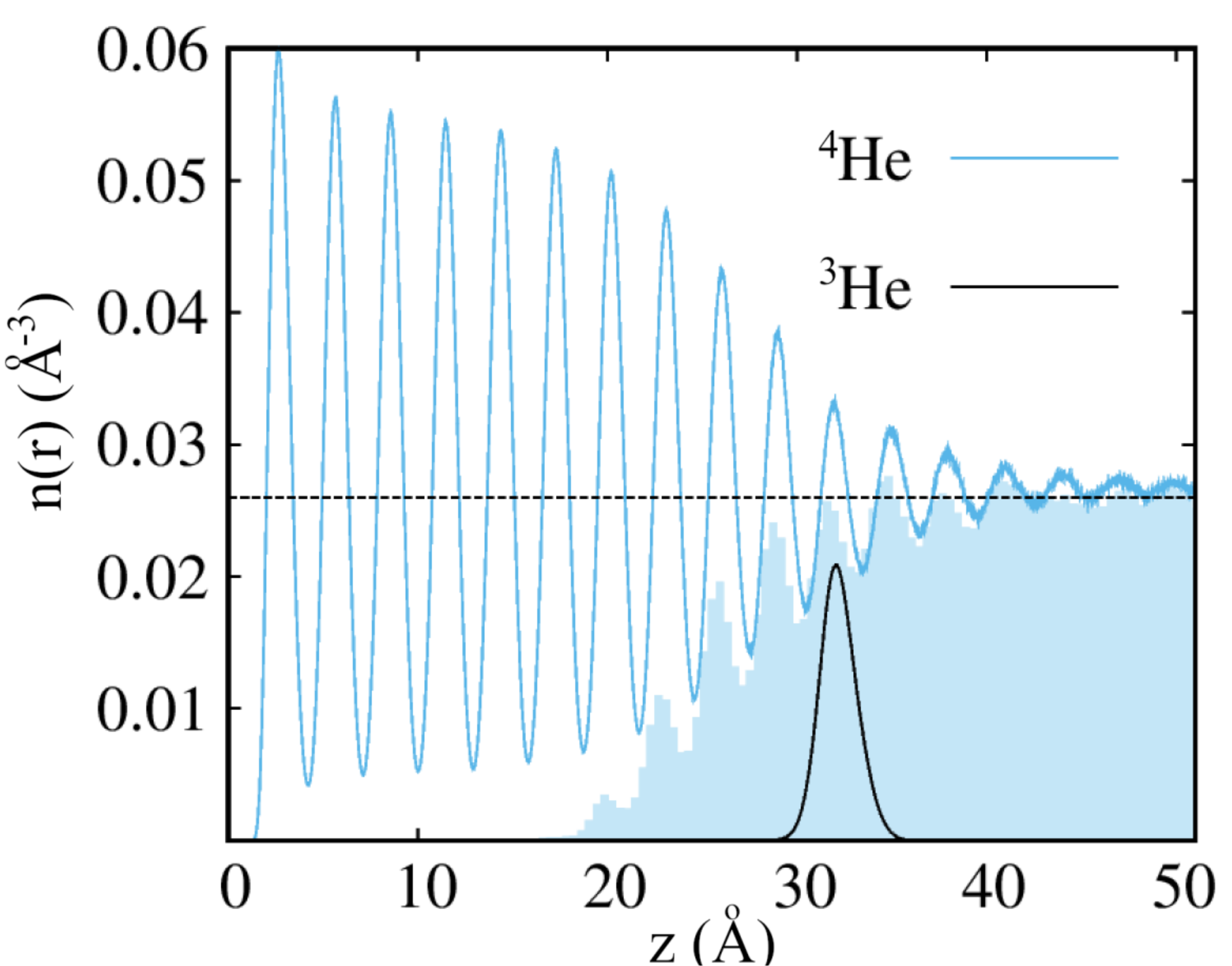}
\caption{Planar averaged density profile of $^4$He in the $z$-direction, perpendicular to the interface between solid (left side) and superfluid (right side) phases, computed at $T=1$ K. Solid black line represents the probability density of position of the single $^3$He atom, in arbitrary units. Shaded part shows the local probability for a $^4$He atom to be part of a quantum-mechanical exchange cycle; units are arbitrary. Dashed line shows the $^4$He melting density $\rho_m$.
}
\label{poh}
\end{figure}
\begin{figure}[h]
\centering
\includegraphics[width=\linewidth]{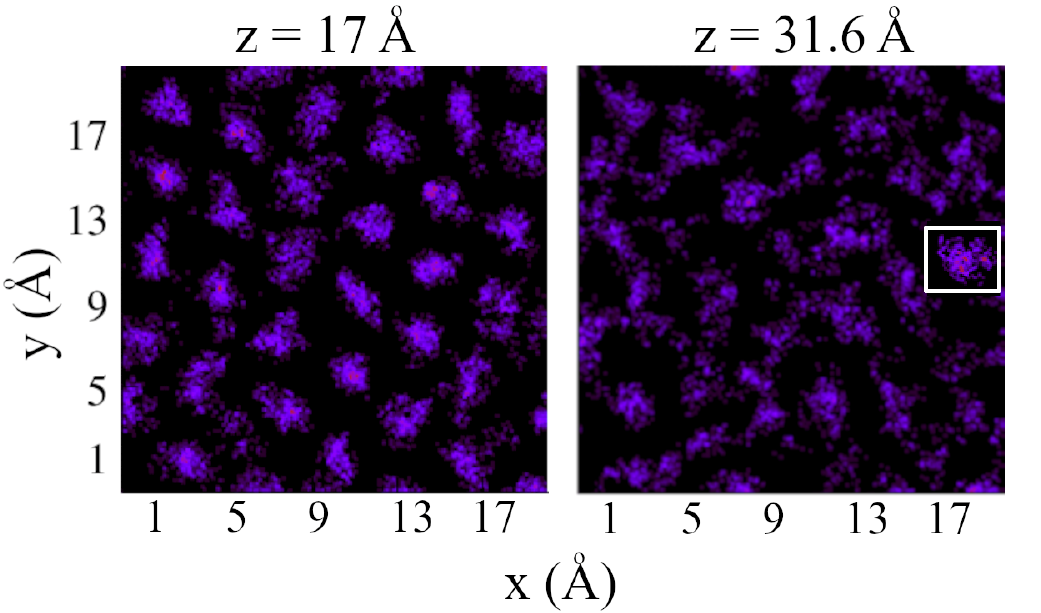}
\caption{Density maps of $^4$He atoms for two layers, located at $z=17$ \AA\ (left) and $z=31.6$ \AA\ (right), with reference to Fig. \ref{poh}. The small rectangular box in the right panel encloses the $^3$He atom}
\label{visual}
\end{figure}

All of the results presented here are obtained at temperature $T=1$ K.
We begin by discussing the first physical scenario investigated here, namely that of the interface between coexisting solid and superfluid phases of $^4$He. The main result is shown in Fig. \ref{poh}, in which the density profile of $^4$He (in \AA$^{-3}$) in the $z$-direction, i.e., perpendicular to the interface, is shown ($z=0$ is the location of the A substrate, i.e., solid helium). The oscillatory structure is consistent with the presence of well-defined solid layers at short distances from the substrate; on moving away from it the oscillations decay, and for $z \gtrsim 40$ \AA, only small ripples persist around a constant value which, within the uncertainty of the calculation, is equal to the freezing density $\rho_f$. We may compare this density profile with the most recently calculated one for this system, using DFT at $T=0$ \cite{ancilotto2005}.  There is reasonable quantitative agreement between the two calculations, as far as the part near the solid substrate is concerned; however, the density profile obtained here shows a transition from crystal to fluid that occurs more gradually, and over a considerably longer distance than in the DFT result.
\\ \indent
As stated in Sec. \ref{meth}, insight into the emergence of superfluid order as one moves away from the solid substrate is offered by the computed frequency with which $^4$He atoms are part of exchange cycles, as a function of their distance from the solid substrate. This quantity is shown (in arbitrary units) by the shaded area in Fig. \ref{poh}. One can see that exchanges begin to appear at $z\approx 20$ \AA, and build up through a few successive adjacent layers which display a physical character intermediate between crystalline and superfluid. We come back to this point below. As one moves further away from the substrate, the probability for a $^4$He atom to be involved in a quantum-mechanical exchanges saturates to a constant value, which is characteristic of the superfluid phase at the freezing density, at $T=1$ K.
\\ \indent
The dark solid line in Fig. \ref{poh}  shows the probability density of position of the single $^3$He atom (in arbitrary units), illustrating how it is located in one the interfacial transition layers mentioned above, positioned at $\sim 31.5$ \AA\ from the substrate. Comparing this result with the original calculation of Ref. \cite {treiner1993} (Fig. 4 therein), based on essentially the same model of the system adopted here, one observes that, aside from the obvious remark that it fails to reproduce the layered structure shown in Fig. \ref{poh}, it also yields a highly localized $^3$He bound state, residing in the only well-defined $^4$He layer yielded by that DFT approach, a layer which is adjacent to the solid substrate. So, there is a degree of qualitative agreement between the two approaches.
\\ \indent
Quantitatively, on the other hand, there are significant differences, chiefly the fact that, as stated above, the calculation carried out here provides a more detailed picture of the interfacial region, and therefore insight into the physical mechanism that determines the location of the $^3$He bound state in a specific interfacial layer. As shown in Fig. \ref{poh}, the density in the layer where the $^3$He atom is located peaks at approximately 0.034 \AA$^{-3}$, which is 15\% lower than the $\sim 0.04$ \AA$^{-3}$ of Ref. \cite{treiner1993}, but still well in the solid part of the $^4$He equilibrium phase diagram. Of course, the layer is a quasi-2D structure, and therefore it makes sense to estimate the effective 2D density of the layer. 
On integrating the $^4$He density profile of Fig. \ref{poh} between $z_1=30.5$ \AA\ and $z_2=33.5$ \AA, we estimate the 2D layer density to be close to 0.08 \AA$^{-2}$, i.e., again well in the crystalline region of the $^4$He 2D phase diagram. At the same time, however, the presence of significant atomic exchanges occurring within this layer (as shown by the shaded area in Fig. \ref{poh}) indicates that the layer cannot be considered a ``crystalline'' solid\footnote{Indeed, quantum-mechanical exchanges have been shown to impart remarkable resilience to overpressurized phases of superfluid $^4$He. See, for instance, Ref. \cite{boninsegni2012b}.}. One possibility is that exchanges may be taking places primarily among atoms in different layers, as opposed to within the layer.
\\ \indent
More intuitive, albeit mainly qualitative insight, can be gained through the direct  visual inspection of  configurations (i.e., many-particle world lines) generated by the Monte Carlo simulation. Two examples are shown in Fig. \ref{visual}, which displays density maps for two $^4$He layers at different distances from wall A (see Fig. \ref{poh} for reference). At $z=17$ \AA\ (left side), the arrangement of $^4$He atoms on a regular (triangular) lattice is clear. This is as close as a simulated system of the size and geometry chosen in this study can approach the actual crystalline structure expected in this system, i.e., hexagonal close packed. It should be kept in mind that atoms find this arrangement {\em spontaneously}, i.e., it is not input ``by hand''.
Furthermore, there appears to be very little overlap between the ``clouds'' representing different atoms, pointing to absence of quantum-mechanical exchanges, consistently with what shown in Fig. \ref{poh}.
\\ \indent
The right side of Fig. \ref{visual} shows instead a density snapshot for a layer at distance $z=31.6$ \AA\ from wall A, which is where the $^3$He atom is located (shown in Fig. \ref{visual} inside a rectangular box).
The difference with the left side is  that atoms are not as orderly positioned (even though remnants of local order can still be seen), and some overlap of different clouds is present to indicate that some atomic exchanges take place within the layer. We come back to this point below.
\\ \indent
We compute the effective mass $m^\star$ of the $^3$He atom using the same procedure adopted in Ref. \cite{boninsegni1995}, i.e., by looking at the diffusion of the $^3$He atom in imaginary time, and obtaining $m^\star$ as
\begin{equation}\label{efm}
    \frac{m}{m^\star} = \frac {2\ m\ \langle[{\bf r}(\beta/2)-{\bf r}(0)]^2\rangle}{3\hbar^2\beta}
\end{equation}
where $\beta=1/T$, $\langle ...\rangle$ stands for thermal average, ${\bf r}(\tau)$ is the position of the $^3$He atom along a path, $0\le\tau\le\beta$ and $m$ is the bare $^3$He mass. We obtain $(m^\star/m)=3.28(5)$, significantly greater than the value (2.3) estimated in Ref. \cite{treiner1993}, but reflecting the high density of the local environment experienced by the $^3$He atom\footnote{It should be noted that the definition given by Eq. \ref{efm} assumes spherical symmetry, but in this case most of the effective mass enhancement arises from in-plane confinement.}.
\\ \indent
\begin{figure}[h]
\centering
\includegraphics[width=\linewidth]{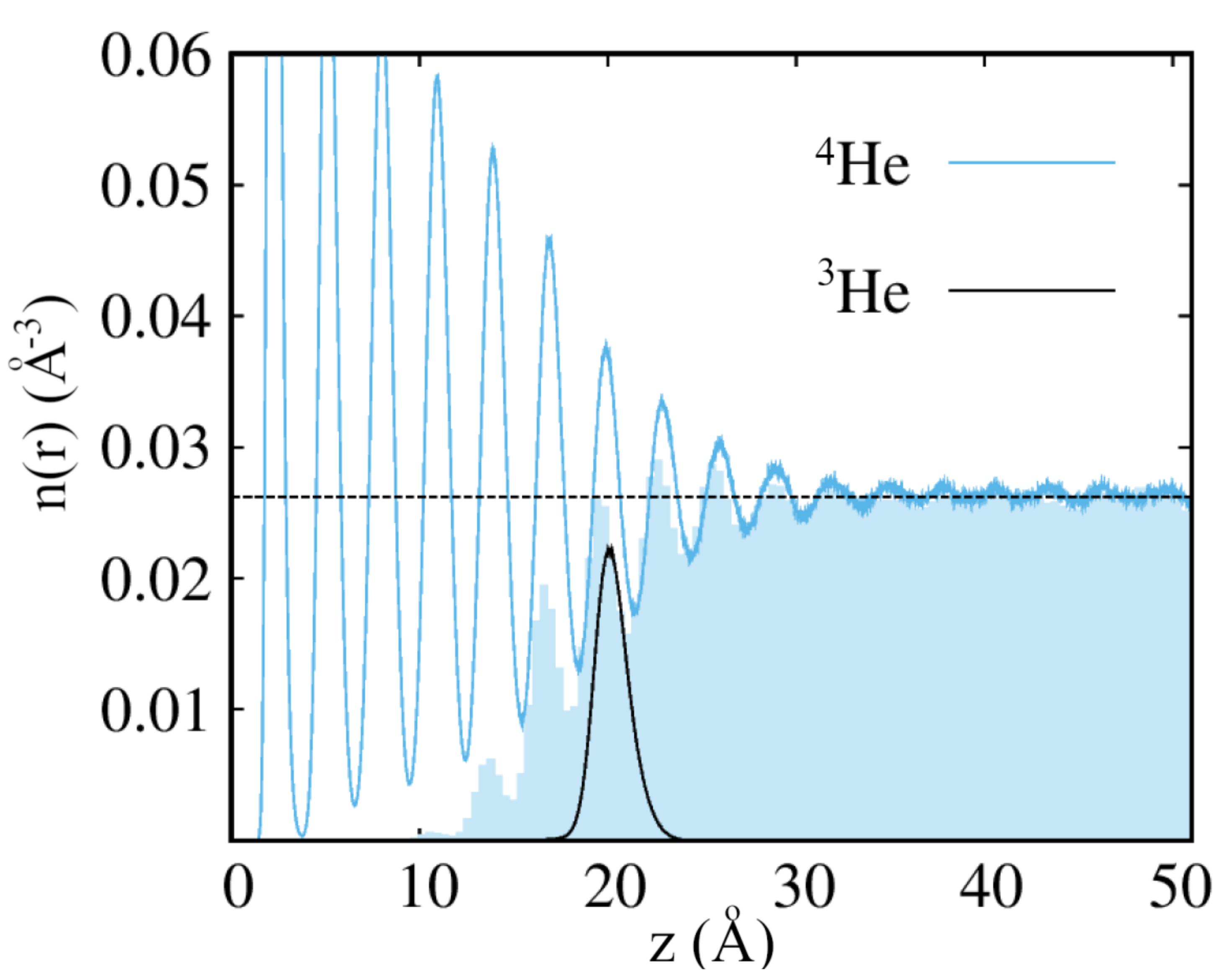}
\caption{Planar averaged density profile of $^4$He in the $z$-direction, perpendicular to the interface between a glass substrate (left side) and superfluid (right side) phases, computed at $T=1$ K. Solid black line represents the probability density of position of the single $^3$He atom, in arbitrary units. Shaded part shows the local probability for a $^4$He atom to be part of a quantum-mechanical exchange cycle; units are arbitrary. Dashed line shows the $^4$He melting density $\rho_m$.
}
\label{pog}
\end{figure}
It is now interesting to contrast these results with those obtained on replacing the solid helium substrate with a considerably more attractive one (set 3 in Table \ref{table1}). Fig. \ref{pog} shows the corresponding $^4$He density profile, which differs from that on solid helium primarily by the more pronounced oscillations in the vicinity of the substrate, which is more strongly attractive and can therefore ``pack'' a greater density of $^4$He atoms in its vicinity. Concurrently, the bound state of the $^3$He atom moves closer to the substrate, mainly because of the greater attraction arising from the higher density of $^4$He atoms near the substrate\footnote{The interaction potential between a He atom and the substrate is only worth $\sim 0.15$ K at the distance at which the bound state form.}.
\\ \indent
Many of the same qualitative physical features found in the previous case remain with this different substrate. The bound state of the $^3$He atom is again localized in a well-defined plane in the interfacial region, in this case one of even higher density (3D density close to 0.04 \AA$^{-3}$, 2D density approximately 0.1 \AA$^{-2}$); however, the $^3$He effective mass is only slightly greater than that found in the previous case, i.e., $3.43(5)\ m$. And, just like in the case of a solid helium substrate, the bound state of $^3$He is located in an interfacial layer characterized by a substantial incidence of quantum exchanges, suggesting that while $^4$He atoms may be localized on the plane, nevertheless they may enjoy, through exchanges, relatively high mobility in the direction perpendicular to the interface.

\section{Discussion and Conclusions}\label{conc}
We have revisited in the work a question that is now relatively old, yet in many respects still unsettled, namely the nature of the bound state of a single $^3$He atom at the interface between solid and superfluid $^4$He. We made use of a computational methodology that is accurate and unbiased, and affords a great deal of insight into the nature of the bound state. Obviously, it is not without shortcomings, the most important being the finite size of the simulation cell. While its length  in the direction perpendicular to the interface is probably sufficient to ensure a fair representation of the interface, the relatively small size of the cell in the transverse direction, namely $\sim 20$ \AA, can be expected to suppress long-wavelength fluctuations (capillary waves) of the interface which, especially given the high mobility of the crystal-superfluid interface, could certainly affect the results. \\ \indent
On the one hand, only calculations on systems of greater size can provide a quantitative assessment of the limitations of the calculation carried out here; in the meantime, we note that the transverse section of the cell utilized here can accommodate about thirty atoms, which is the typical size of most QMC simulations of 2D $^4$He (see, for instance, Ref. \cite{gordillo1998}. Also, note for reference that, the only existing comparable calculation (Ref. \cite{khairallah2006}), whose scope is not too dissimilar from that of the present work, made use of a cell of transverse size $\sim 17$ \AA, and its results are deemed quantitatively reliable. 
\\ \indent
Some of the existing theoretical predictions are qualitatively confirmed by the calculation illustrated here; for example, it is found that the $^3$He bound state is localized in one of the quasi-2D interfacial layers, although such a layer may not necessarily be ``non-solid'', as suggested in previous works \cite{treiner1993}; ``non-crystalline'' may be more appropriate, as the relatively high incidence of atomic exchanges in the layer points to a possibly high-density, disordered, anisotropic  ''glassy'' superfluid phase \cite{boninsegni2006} with superflow only in the direction perpendicular to the plane\footnote{The  more intriguing scenario of concomitant in-plane crystalline order and superflow was considered, but is not supported by a direct visual inspection of in-plane atomic configurations, as shown in Fig. \ref{visual}.}.
\\ \indent
Altogether, the evidence of a $^3$He bound state is fairly robust; however, this study cannot add cogent  information about the magnitude of the $^3$He binding energy, for which conflicting estimates exist. Unfortunately, the relatively large number of atoms in the simulation, which is necessary in order to obtain a meaningful representation of the interface, renders the direct calculation of the binding energy numerically unfeasible, as it entails the subtraction of two extensive quantities. Other possible approaches, based on e.g., re-weighting, or the use of the single-particle Matsubara Green function, available within the Worm Algorithm and successfully utilized to compute the activation energy of vacancies and interstitials in solid $^4$He \cite{fate}, or the asymptotic behavior of the $^3$He probability density of position, were not found to afford the numerical precision required in this case. Further work along these lines is in progress.
\\ \indent
One remark that can be made, however, is that the significant difference between the values of the effective mass of the $^3$He atom obtained here and in Ref. \cite{treiner1993}, as well as the very different $^4$He density profiles, raise some doubts about the quantitative accuracy of the binding energy estimates furnished therein.

\section{Declaration of Competing Interests}
The author declares that he has no known competing financial interests or personal relationships that could have appeared to influence the work reported in this paper.

\section{Acknowledgments}
This work was supported by the Natural Sciences and Engineering Research Council of Canada.

 \bibliographystyle{elsarticle-num} 
 \bibliography{refs}

\begin{thebibliography}{10}
\expandafter\ifx\csname url\endcsname\relax
  \def\url#1{\texttt{#1}}\fi
\expandafter\ifx\csname urlprefix\endcsname\relax\def\urlprefix{URL }\fi
\expandafter\ifx\csname href\endcsname\relax
  \def\href#1#2{#2} \def\path#1{#1}\fi

\bibitem{edwards1978}
D.~O. Edwards, W.~F. Saam, The free surface of liquid helium, in: D.~E. Brewer
  (Ed.), Progress in Low Temperature Physics, Vol. {VIIA}, North-Holland,
  Amsterdam, The Netherlands, 1978, pp. 283--369.

\bibitem{bashkin1980}
E.~P. Bashkin, ({He}$^3$)$_2$, {van der Waals} molecular dimers in solutions of
  the quantum liquids {He}$^3$--{He}$_{{\rm II}}$, Sov. Phys. JETP 51~(1)
  (1980) 181--189.

\bibitem{miyake1983}
K.~Miyake, Fermi liquid theory of dilute submonolayer $^3$he on thin $^4$he ii
  film: Dimer bound state and cooper pairs, Prog. Theor. Phys. 69~(6) (1983)
  1794--1797.
\newblock \href {https://doi.org/10.1143/PTP.69.1794}
  {\path{doi:10.1143/PTP.69.1794}}.

\bibitem{andreev1966}
A.~F. Andreev, Surface tension of weak helium isotope solutions, Sov. Phys.
  JETP 23~(5) (1966) 939--941.

\bibitem{atkins1965}
K.~R. Atkins, Y.~Narahara, Surface tension of liquid ${^4\mathrm{He}}$, Phys.
  Rev. 138 (1965) A437--A441.
\newblock \href {https://doi.org/10.1103/PhysRev.138.A437}
  {\path{doi:10.1103/PhysRev.138.A437}}.

\bibitem{lekner1970}
J.~Lekner, Theory of surface states of $^3${He} atoms in liquid $^4${He},
  Philos. Mag. 22~(178) (1970) 669--673.
\newblock \href {https://doi.org/10.1080/14786437008220937}
  {\path{doi:10.1080/14786437008220937}}.

\bibitem{pavloff1991}
N.~Pavloff, J.~Treiner, $^3${He} impurity states on liquid $^4${He}: From thin
  films to the bulk surface, J. Low Temp. Phys. 83~(5-6) (1991) 331--349.
\newblock \href {https://doi.org/10.1007/bf00683631}
  {\path{doi:10.1007/bf00683631}}.

\bibitem{treiner1993}
J.~Treiner, Helium mixtures on weak binding substrates, J. Low Temp. Phys.
  92~(1-2) (1993) 1--9.
\newblock \href {https://doi.org/10.1007/BF00681869}
  {\path{doi:10.1007/BF00681869}}.

\bibitem{landau1948}
L.~Landau, I.~Pomeranchuk, {On the Movement of Foreign Particles in Helium II},
  Dokl. Akad. Nauk SSSR 59~(4) (1948) 669--670.

\bibitem{carmi1988}
Y.~Carmi, E.~Polturak, S.~G. Lipson, Roughening transition in dilute
  $^3${He}-$^4${He} mixture crystals, Phys. Rev. Lett. 62~(12) (1989)
  1364--1367.
\newblock \href {https://doi.org/10.1103/PhysRevLett.62.1364}
  {\path{doi:10.1103/PhysRevLett.62.1364}}.

\bibitem{wang1992}
C.~L. Wang, G.~Agnolet, Effects of $^3${He} impurities on the $^4${He}
  solid-liquid interface, J. Low Temp. Phys. 89~(3-4) (1992) 759--762.
\newblock \href {https://doi.org/10.1007/BF00694135}
  {\path{doi:10.1007/BF00694135}}.

\bibitem{ketola1993}
K.~S. Ketola, R.~B. Hallock, Effect of $^{3}\mathrm{He}$ on the wetting of
  $^{4}\mathrm{He}$ to a cesium-coated substrate, Phys. Rev. Lett. 71~(20)
  (1993) 3295--3298.
\newblock \href {https://doi.org/10.1103/PhysRevLett.71.3295}
  {\path{doi:10.1103/PhysRevLett.71.3295}}.

\bibitem{draisma1994}
W.~Draisma, M.~Eggenkamp, P.~Pinkse, R.~Jochemsen, G.~Frossati, Possible
  observation of the substrate state in $^3${He}-$^4${He} mixture films,
  Physica B 194-196 (1994) 853--854.
\newblock \href {https://doi.org/10.1016/0921-4526(94)90756-0}
  {\path{doi:10.1016/0921-4526(94)90756-0}}.

\bibitem{sheldon1995}
P.~A. Sheldon, R.~B. Hallock, Absence of a substrate state for $^3${He} in a
  $^3${He}-$^4${He} bulk mixture in proximity to a strong-binding surface,
  Phys. Rev. B 52~(17) (1995) 12530--12533.
\newblock \href {https://doi.org/10.1103/PhysRevB.52.12530}
  {\path{doi:10.1103/PhysRevB.52.12530}}.

\bibitem{ross1995}
D.~Ross, P.~Taborek, J.~E. Rutledge, Bound states of $^3${He} at the
  helium-cesium interface, Phys. Rev. Lett. 74~(22) (1995) 4483--4486.
\newblock \href {https://doi.org/10.1103/PhysRevLett.74.4483}
  {\path{doi:10.1103/PhysRevLett.74.4483}}.

\bibitem{rolley1995}
E.~Rolley, S.~Balibar, C.~Guthmann, P.~Nozi\`eres, Adsorption of $^3${He} on
  $^4${He} crystal surfaces, Physica 210~(3-4) (1995) 397--402.
\newblock \href {https://doi.org/10.1016/0921-4526(94)01126-L}
  {\path{doi:10.1016/0921-4526(94)01126-L}}.

\bibitem{chang2021}
Z.~G. Cheng, J.~Beamish, In situ monitoring distribution and migration of
  $^3${He} in liquid-solid $^4${He} mixtures, Phys. Rev. Res. 3~(2) (2021).
\newblock \href {https://doi.org/10.1103/physrevresearch.3.023136}
  {\path{doi:10.1103/physrevresearch.3.023136}}.

\bibitem{khairallah2006}
S.~A. Khairallah, D.~M. Ceperley, Superfluidity of dense $^{4}\mathrm{He}$ in
  vycor, Phys. Rev. Lett. 95~(18) (2005) 185301.
\newblock \href {https://doi.org/10.1103/PhysRevLett.95.185301}
  {\path{doi:10.1103/PhysRevLett.95.185301}}.

\bibitem{pricaupenko1995}
L.~Pricaupenko, J.~Treiner, Phase separation of liquid $^3${He}-$^4${He}
  mixtures: Effect of confinement, Phys. Rev. Lett. 74~(3) (1995) 430--433.
\newblock \href {https://doi.org/10.1103/PhysRevLett.74.430}
  {\path{doi:10.1103/PhysRevLett.74.430}}.

\bibitem{boninsegni2010}
M.~Boninsegni, Thin helium film on a glass substrate, J. Low Temp. Phys.
  159~(3-4) (2010) 441--451.
\newblock \href {https://doi.org/10.1007/s10909-009-0143-1}
  {\path{doi:10.1007/s10909-009-0143-1}}.

\bibitem{grilly1973}
E.~R. Grilly, Pressure-volume-temperature relations in liquid and solid
  $^4${He}, J. Low Temp. Phys. 11~(1-2) (1973) 33--52.
\newblock \href {https://doi.org/10.1007/bf00655035}
  {\path{doi:10.1007/bf00655035}}.

\bibitem{aziz1979}
R.~A. Aziz, V.~P.~S. Nain, J.~S. Carley, W.~L. Taylor, G.~T. McConville, An
  accurate intermolecular potential for helium, J. Chem. Phys. 70~(9) (1979)
  4330--4342.
\newblock \href {https://doi.org/10.1063/1.438007}
  {\path{doi:10.1063/1.438007}}.

\bibitem{mezz}
F.~Mezzacapo, M.~Boninsegni, Superfluidity and quantum melting of {\em
  p}-{H}$_2$ clusters, Phys. Rev. Lett. 97~(4) (2006) 045301.
\newblock \href {https://doi.org/10.1103/PhysRevLett.97.045301}
  {\path{doi:10.1103/PhysRevLett.97.045301}}.

\bibitem{mezz2}
F.~Mezzacapo, M.~Boninsegni, Structure, superfluidity, and quantum melting of
  hydrogen clusters, Phys. Rev. A 75~(3) (2007) 033201.
\newblock \href {https://doi.org/10.1103/PhysRevA.75.033201}
  {\path{doi:10.1103/PhysRevA.75.033201}}.

\bibitem{worm}
M.~Boninsegni, N.~Prokof'ev, B.~Svistunov, {Worm Algorithm for Continuous-Space
  Path Integral Monte Carlo Simulations}, Phys. Rev. Lett. 96~(7) (2006)
  070601.
\newblock \href {https://doi.org/10.1103/PhysRevLett.96.070601}
  {\path{doi:10.1103/PhysRevLett.96.070601}}.

\bibitem{worm2}
M.~Boninsegni, N.~V. Prokof'ev, B.~V. Svistunov, Worm algorithm and
  diagrammatic {Monte Carlo}: A new approach to continuous-space path integral
  {Monte Carlo} simulations, Phys. Rev. E 74~(3) (2006) 036701.
\newblock \href {https://doi.org/10.1103/PhysRevE.74.036701}
  {\path{doi:10.1103/PhysRevE.74.036701}}.

\bibitem{rmp95}
D.~M. Ceperley, Path integrals in the theory of condensed helium, Rev. Mod.
  Phys. 67~(2) (1995) 279--355.
\newblock \href {https://doi.org/10.1103/RevModPhys.67.279}
  {\path{doi:10.1103/RevModPhys.67.279}}.

\bibitem{psb}
M.~Boninsegni, S.~Moroni, {Population size bias in diffusion Monte Carlo},
  Phys. Rev. E 86~(5) (2012) 056712.
\newblock \href {https://doi.org/10.1103/PhysRevE.86.056712}
  {\path{doi:10.1103/PhysRevE.86.056712}}.

\bibitem{phasesep}
M.~Boninsegni, {Phase Separation in Mixtures of Hard Core Bosons}, Phys. Rev.
  Lett. 87~(8) (2001) 087201.
\newblock \href {https://doi.org/10.1103/PhysRevLett.87.087201}
  {\path{doi:10.1103/PhysRevLett.87.087201}}.

\bibitem{jltp2}
M.~Boninsegni, Permutation sampling in path integral {Monte Carlo}, J. Low
  Temp. Phys. 141~(1-2) (2005) 27--46.
\newblock \href {https://doi.org/10.1007/s10909-005-7513-0}
  {\path{doi:10.1007/s10909-005-7513-0}}.

\bibitem{paesani}
Y.~Kwon, F.~Paesani, K.~B. Whaley, Local superfluidity in inhomogeneous quantum
  fluids, Phys. Rev. B 74 (2006) 174522.
\newblock \href {https://doi.org/10.1103/PhysRevB.74.174522}
  {\path{doi:10.1103/PhysRevB.74.174522}}.

\bibitem{mezz3}
F.~Mezzacapo, M.~Boninsegni, Local superfluidity of parahydrogen clusters,
  Phys. Rev. Lett. 100 (2008) 145301.
\newblock \href {https://doi.org/10.1103/PhysRevLett.100.145301}
  {\path{doi:10.1103/PhysRevLett.100.145301}}.

\bibitem{dang2008}
L.~Dang, M.~Boninsegni, L.~Pollet, Disorder-induced superfluidity, Phys. Rev. B
  79~(21) (2009) 214529--214529.
\newblock \href {https://doi.org/10.1103/PhysRevB.79.214529}
  {\path{doi:10.1103/PhysRevB.79.214529}}.

\bibitem{boninsegni2006}
M.~Boninsegni, N.~Prokof'ev, B.~Svistunov, Superglass phase of
  $^{4}\mathrm{He}$, Phys. Rev. Lett. 96~(10) (2006) 105301.
\newblock \href {https://doi.org/10.1103/PhysRevLett.96.105301}
  {\path{doi:10.1103/PhysRevLett.96.105301}}.

\bibitem{boninsegni1995}
M.~Boninsegni, D.~M. Ceperley, Path integral monte carlo simulation of isotopic
  liquid helium mixtures, Phys. Rev. Lett. 74~(12) (1995) 2288--2291.
\newblock \href {https://doi.org/10.1103/PhysRevLett.74.2288}
  {\path{doi:10.1103/PhysRevLett.74.2288}}.

\bibitem{ancilotto2005}
F.~Ancilotto, M.~Barranco, F.~Caupin, R.~Mayol, M.~Pi, Freezing of $^4${He} and
  its liquid-solid interface from density functional theory, Phys. Rev. B
  72~(21) (2005).
\newblock \href {https://doi.org/10.1103/physrevb.72.214522}
  {\path{doi:10.1103/physrevb.72.214522}}.

\bibitem{boninsegni2012b}
M.~Boninsegni, L.~Pollet, N.~Prokof’ev, B.~Svistunov, Role of bose statistics
  in crystallization and quantum jamming, Phys. Rev. Lett. 109~(2) (2012)
  25302--25302.
\newblock \href {https://doi.org/10.1103/PhysRevLett.109.025302}
  {\path{doi:10.1103/PhysRevLett.109.025302}}.

\bibitem{gordillo1998}
M.~C. Gordillo, D.~M. Ceperley, Path-integral calculation of the
  two-dimensional helium-four phase diagram, Phys. Rev. B 58~(10) (1998)
  6447--6454.
\newblock \href {https://doi.org/10.1103/PhysRevB.58.6447}
  {\path{doi:10.1103/PhysRevB.58.6447}}.

\bibitem{fate}
M.~Boninsegni, A.~Kuklov, L.~Pollet, N.~Prokof’ev, B.~Svistunov, M.~Troyer,
  Fate of vacancy-induced supersolidity in $^4${He}, Phys. Rev. Lett. 97~(8)
  (2006) 080401--080401.
\newblock \href {https://doi.org/10.1103/PhysRevLett.97.080401}
  {\path{doi:10.1103/PhysRevLett.97.080401}}.

\end{thebibliography}

%% else use the following coding to input the bibitems directly in the
%% TeX file.

% \begin{thebibliography}{00}

% %% \bibitem{label}
% %% Text of bibliographic item

% \bibitem{}

% \end{thebibliography}
\end{document}